\documentclass[codesnippet]{jss1}
\usepackage{multirow}
\usepackage{longtable}
\usepackage{ltcaption}
\usepackage{listings}
\usepackage{graphicx}
\usepackage{amsmath}
\usepackage{natbib}
\usepackage{color}
\usepackage{amssymb}

\usepackage[singlelinecheck=off]{caption}

\author{Victoria N Nyaga \\Institute of Public Health, \\ Hasselt University \And 
        Marc Arbyn\\Institute of Public Health \And
        Marc Aerts\\Hasselt University}
\title{\pkg{CopulaDTA}: An \proglang{R} Package for Copula Based Bivariate Beta-Binomial Models for Diagnostic Test Accuracy Studies in a Bayesian Framework}

\Plainauthor{Victoria Nyaga, Marc Arbyn, Marc Aerts} 
\Plaintitle{CopulaDTA: An R Package for Copula Based Bivariate Beta-Binomial Model for Diagnostic Test Accuracy Studies in Bayesian Framework} 
\Shorttitle{\pkg{CopulaDTA}: Bivariate Beta-Binomial Models in \proglang{R}} 

\Abstract{
The current statistical procedures implemented in statistical software packages for pooling of diagnostic test accuracy data include hSROC regression \citep{hsroc} and the bivariate random-effects meta-analysis model (BRMA) (\citet{Reitsma}, \citet{Arends}, \citet{Chu}, \citet{Rileya}). However, these models do not report the overall mean but rather the mean for a central study with random-effect equal to zero and have difficulties estimating the correlation between sensitivity and specificity when the number of studies in the meta-analysis is small and/or when the between-study variance is relatively large \citep{Rileyb}.

This tutorial on advanced statistical methods for meta-analysis of diagnostic accuracy studies discusses and demonstrates Bayesian modeling using \pkg{CopulaDTA}~\citep{copuladta} package in \proglang{R}~\citep{R} to fit different models to obtain the meta-analytic parameter estimates. The focus is on the joint modelling of sensitivity and specificity using copula based bivariate beta distribution. Essentially, we extend the work of \citet{Nikolo} by: \textbf{i}) presenting the Bayesian approach which offers flexibility and ability to perform complex statistical modelling even with small data sets and \textbf{ii}) including covariate information, and \textbf{iii}) providing an easy to use code. The statistical methods are illustrated by re-analysing data of two published meta-analyses.

Modelling sensitivity and specificity using the bivariate beta distribution provides marginal as well as study-specific parameter estimates as opposed to using bivariate normal distribution (e.g., in BRMA) which only yields study-specific parameter estimates. Moreover, copula based models offer greater flexibility in modelling different correlation structures in contrast to the normal distribution which allows for only one correlation structure.
}
\Plainkeywords{diagnostic test accuracy, meta-analysis, Bayesian, random-effects, copula, R} 

\Address{
	Victoria N Nyaga\\
	Unit of Cancer Epidemiology\\
	Scientific Institute of Public Health\\
	Juliette Wytsmanstraat 14\\
	1050 Brussels, Belgium\\
	E-mail: \email{victoria.nyawiranyaga@wiv-isp.be} \\
	E-mail: \email{victoria.nyaga@uhasselt.be}

	Marc Arbyn\\
	Unit of Cancer Epidemiology\\
	Scientific Institute of Public Health\\
	Juliette Wytsmanstraat 14\\
	1050 Brussels, Belgium\\
	E-mail: \email{marc.arbyn@wiv-isp.be}
	
    Marc Aerts\\
	Center for Statistics\\
	Hasselt University\\
	Martelarenlaan 42\\
	3500 Hasselt, Belgium\\
	E-mail: \email{marc.aerts@uhasselt.be}
	}

\begin{document}


\section[Introduction]{Introduction} \label{1}
In a systematic review of diagnostic test accuracy, the statistical analysis section aims at estimating the average (across studies) sensitivity and specificity of a test and the variability thereof, among other measures. There tends to be a negative correlation between sensitivity and specificity, which postulates the need for correlated data models. The analysis is statistically challenging because the user \textbf{i}) deals with two summary statistics, \textbf{ii}) has to account for correlation between sensitivity and specificity, \textbf{iii}) has to account for heterogeneity in sensitivity and specificity across the studies and \textbf{iv}) should be allowed to incorporate covariates. 

Currently, the  HSROC regression \citep{hsroc} or the bivariate random-effects meta-analysis model (BRMA) (\citet{Reitsma}, \citet{Arends}, \citet{Chu}) are recommended for pooling of diagnostic test accuracy data. These models fit a bivariate normal distribution which allows for only one correlation structure to the logit transformed sensitivity and specificity. The resulting distribution has no closed form and therefore the mean sensitivity and specificity is only estimated after numerical integration or other approximation methods, an extra step which is rarely taken.

Within the maximum likelihood estimation methods, the BRMA and HSROC models have difficulties in convergence and estimating the correlation parameter when the number of studies in the meta-analysis are small and/or when the between-study variances are relatively large \citep{Takwoingi}.  When the correlation is close to the boundary of its parameter space, the between study variance estimates from the BRMA are upwardly biased as they are inflated to compensate for the range restriction on the correlation parameter \citep{Rileyb}. According to \citet{Rileya} this occurs because the maximum likelihood estimator truncates the between-study covariance matrix on the boundary of its parameter space, and this often occurs when the within-study variation is relatively large or the number of studies is small.

The BRMA and HSROC assume that the transformed data is approximately normal with constant variance, however for sensitivity and specificity and proportions in generals, the mean and variance depend on the underlying probability. Therefore, any factor affecting the probability will change the mean and the variance. This implies that the in models where the predictors affect the mean but assume a constant variance will not be adequate. 

Joint modelling of study specific sensitivity and specificity using existing or copula based bivariate beta distributions overcomes the above mentioned difficulties. Since both sensitivity and specificity take values in the interval space (0, 1), it is a more natural choice to use a beta distribution to describe their distribution across studies, without the need for any transformation. The beta distribution is conjugate to the binomial distribution and therefore it is easy to integrate out the random-effects analytically giving rise to the beta-binomial marginal distributions. Moreover no further integration is needed to obtain the meta-analytically pooled sensitivity and specificity. Previously, \citet{Cong} fitted separate beta-binomial models to the number of true positives and the number of false positives. While the model ignores correlation between sensitivity and specificity, \cite{Cong} reported that the model estimates are comparable to those from the SROC model \citep{Moses}, the predecessor of the HSROC model. 

According to \citet{Rileyc}, ignoring the correlation would have negligible influence on the meta-analysis results when the within-study variability is large relative to the between-study variability. By utilising the correlation, we allow borrowing strength across sensitivities and specificities resulting in smaller standard errors. The use of copula based mixed models within the frequentist framework for meta-analysis of diagnostic test accuracy was recently introduced by \cite{Nikolo} who evaluated the joint density numerically.  

This tutorial, presents and demonstrates hierarchical mixed models for meta-analysis of diagnostic accuracy studies. In the first level of the hierarchy, given sensitivity and specificity for each study, two binomial distributions are used to describe the variation in the number of true positives and true negatives among the diseased and healthy individuals, respectively. In the second level, we  model the unobserved sensitivities and specificities using a bivariate distribution. While hierarchical models are used, the focus of meta-analysis is on the pooled average across studies and rarely on a given study estimate.

The methods are demonstrated using datasets from two previously published meta-analyses: \textbf{a}) on diagnostic accuracy of telomerase in urine as a tumour marker for the diagnosis of primary bladder cancer from \citet{Glas} previously used by \citet{Rileya} and \citet{Nikolo} since it is a problematic dataset that has convergence issues caused by the correlation parameter being estimated to be -1 and has no covariate and \textbf{b}) on the comparison of the sensitivity and specificity of human papillomavirus testing (using the HC2 assay) versus repeat cytology to triage women with minor cytological cervical lesions to detect underlying cervical precancer from \citet{Arbyn}. the second dataset is used to demonstrate meta-regression with one covariate which can be naturally extended to include several covariates.

The layout of this tutorial is as follows: in Section~\ref{2} introduces the concept of copula theory and different bivariate distributions for sensitivity and specificity. The software implementation and model selection in a Bayesian framework is discussed in Section~\ref{3}. The two aforementioned datasets are introduced in Section~\ref{4}. Application of software, code examples and the results of the models fitted to the data are presented in Section~\ref{5} and~\ref{6}. A brief discussion is found in Section~\ref{7} and a conclusion in Section~\ref{8}.

\section[Methods]{Statistical methods for meta-analysis} \label{2}
\subsection{Definition of copula function}
A bivariate copula function describes the dependence structure between two random variables. Two random variables $X_1$ and $X_2$ are joined by a copula function C if their joint cumulative distribution function can be written as 
\begin{equation}\label{eq:1}
F(x_1,~x_2 ) = C(F_1 (x_1), ~ F_2 (x_2 )),~ -\infty ~\le ~ x_1,~x_2~ \le~+\infty.
\end{equation}

According to the theorem of \citet{Skylar}, there exists for every bivariate (multivariate in extension) distribution a copula representation C which is unique for continuous random variables. If the joint cumulative distribution function and the two marginals are known, then the copula function can be written as
\begin{equation}\label{eq:2}
C(u, ~v) = F(F_1^{-1} (u), ~F_2^{-1} (v)),~ 0 \le~ u, ~v ~\le~ 1. 
\end{equation}
A 2-dimensional copula is in fact simply a 2-dimensional cumulative function restricted to the unit square with standard uniform marginals. A comprehensive overview of copulas and their mathematical properties can be found in \citet{Nelsen}. 
To obtain the joint probability density, the joint cumulative distribution in Equation~\ref{eq:1} should be differentiated to yield
\begin{equation}\label{eq:3}
f(x_1, ~x_2) = f_1(x_1) ~f_2(x_2 ) ~c(F_1(x_1), ~F_2 (x_2)),
\end{equation}
where $f_1$ and $f_2$ denote the marginal density functions and c the copula density function corresponding to the copula cumulative distribution function C. Therefore from Equation~\ref{eq:3}, a bivariate probability density can be expressed using the marginal and the copula density, given that the copula function is absolutely continuous and twice differentiable.

When the functional form of the marginal and the joint densities are known, the copula density can be derived as follows
\begin{equation}\label{eq:4}
c(F_1(x_1), ~F_2(x_2)) = \frac{f(x_1, ~x_2)}{f_1 (x_1 )~ f_2 (x_2 )} .
\end{equation}								

While our interest does not lie in finding the copula function, Equation~\ref{eq:3} and~\ref{eq:4} serve to show how one can move from the copula function to the bivariate density or vice-versa, given that the marginal densities are known. The decompositions allow for constructions of other and possible better models for the variables than would be possible if we limited ourselves to only existing standard bivariate distributions. 

We finish this section by mentioning an important implication when Sklar's theorem is extended to a meta-regression setting with covariates.  According to \citet{Patton}, it is important that the conditioning variable remains the same for both marginal distributions and the copula, as otherwise the joint distribution might not be properly defined. This implies that covariate information should be introduced in both the marginals and the association parameters of the model.

\subsection{The hierarchical model}
Since there are two sources of heterogeneity in the data, the within- and between-study variability, the parameters involved in a meta-analysis of diagnostic accuracy studies vary at two levels. For each study \textit{i}, \textit{i = 1, ..., n}, let $Y_{i}~=~(Y_{i1},~ Y_{i2})$  denote the true positives and true negatives, $N_{i}~=~( N_{i1},~ N_{i2})$ the diseased and healthy individuals respectively, and $\pi_{i}~ =~ (\pi_{i1},~ \pi_{i2})$ represent the `unobserved' sensitivity and specificity respectively.  

Given study-specific sensitivity and specificity, two separate binomial distributions describe the distribution of true positives and true negatives among the diseased and the healthy individuals as follows
\begin{equation}\label{eq:5}
Y_{ij}~ |~ \pi_{ij}, ~\textbf{x}_i~ \sim~ bin(\pi_{ij},~ N_{ij}), i~=~1, ~\dots ~n, ~j~=~1, ~2,
\end{equation} 						
where $\textbf{x}_i$ generically denotes one or more covariates, possibly affecting $\pi_{ij}$. Equation ~\ref{eq:5} forms the higher level of the hierarchy and models the within-study variability. The second level of the hierarchy aims to model the between study variability of sensitivity and specificity while accounting for the inherent negative correlation thereof, with a bivariate distribution as follows
\begin{equation}\label{eq:6}
\begin{pmatrix}
	g(\pi_{i1})\\
	g(\pi_{i2}) 
\end{pmatrix} \sim f(g(\pi_{i1}),~ g(\pi_{i2}))~ =~ f(g(\pi_{i1})) ~f(g(\pi_{i2})) ~c(F_1(g(\pi_{i1})),~ F_2(g(\pi_{i2}))),
\end{equation}		
where \textit{g}(.) denotes a transformation that might be used to modify the (0, 1) range to the whole real line. While it is critical to ensure that the studies included in the meta-analysis satisfy the specified entry criterion, there are study specific characteristics like different test thresholds and other unobserved differences that give rise to the second source of variability, the between-study variability. It is indeed the difference in the test thresholds between the studies that gives rise to the correlation between sensitivity and specificity. Including study level covariates allows us to model part of the between-study variability. The covariate information can and should \citep{Patton} be used to model the mean as well as the correlation between sensitivity and specificity.

In the next section we give more details on different bivariate distributions $f(g(\pi_{i1}),~g(\pi_{i2}))$ constructed using the logit or identity link function \textit{g}(.), different marginal densities and/or different copula densities \textit{c}. We discuss their implications and demonstrate their application in meta-analysis of diagnostic accuracy studies.  An overview of suitable parametric families of copula for mixed models for diagnostic test accuracy studies was recently given by \citet{Nikolo}. Here, we consider five copula functions which can be plugged in Equation~\ref{eq:3} to model negative correlation.

\subsubsection{Bivariate Gaussian copula}
Given the density and the distribution function of the univariate and bivariate standard normal distribution with correlation parameter $\rho \in (-1, 1)$, the bivariate Gaussian copula function and density is expressed \citep{Meyer} as
\begin{align}
C(u, ~v, ~\rho) ~&=~ \Phi_2(\Phi^{-1}(u),~ \Phi^{-1}(v),~ \rho), \nonumber\\
c(u, ~v, ~\rho) ~&=~  \frac{1}{\sqrt{1~-~\rho^2}}  ~exp\bigg(\frac{2~\rho~\Phi^{-1}(u) ~\Phi^{-1}(v) - \rho^2~ (\Phi^{-1}(u)^2 + \Phi^{-1}(v)^2)}{2~(1 - \rho^2)}\bigg ). \label{eq:7}
\end{align}					

The logit transformation is often used in binary logistic regression to relate the probability of ``success" (coded as 1, failure as 0) of the binary response variable with the linear predictor model that theoretically can take values over the whole real line. In diagnostic test accuracy studies, the `unobserved' sensitivities and specificities can range from 0 to 1 whereas their logits = $log⁡(\frac{\pi_{ij}}{1 ~-~ \pi_{ij}})$ can take any real value allowing to use the normal distribution as follows 
\begin{equation}
logit (\pi_{ij}) ~\sim~ N(\mu_j, ~\sigma_j) ~<=> ~logit (\pi_{ij}) ~=~ \mu_j ~+~ \varepsilon_{ij},
\label{eq:8}
\end{equation}
where, $\mu_j$ is a vector of the mean sensitivity and specificity for a study with zero random effects, and $\varepsilon_{i}$ is a vector of random effects associated with study \textit{i}.
Now \textit{u} is the normal distribution function of $logit(\pi_{i1}$) with parameters $\mu_1$ and $\sigma_1$, \textit{v} is the normal distribution function of $logit(\pi_{i2})$ with parameters $\mu_2$ and $\sigma_2$,  $\Phi_2$ is the distribution function of a bivariate standard normal distribution with correlation parameter $\rho \in (-1, ~1)$ and $\Phi^{-1}$  is the quantile of the standard normal distribution. In terms of $\rho$, Kendall's tau is expressed as ($\frac{2}{\pi}$)arcsin$(\rho)$. 

With simple algebra the copula density in Equation~\ref{eq:7} with normal marginal distributions simplifies to 
\begin{equation}
c(u, ~v, ~\rho) = \frac{1}{\sqrt{1 - \rho^2}} ~exp\bigg(\frac{1}{2 ~(1 ~-~ \rho^2)}~ \bigg( \frac{2~\rho~(x - \mu_1)~(y - \mu_2)}{\sigma_1 ~\sigma_2} ~-~ \rho^2 ~\bigg (\frac{{x ~-~ \mu_1}^2}{\sigma_1} ~+~ \frac{{y ~-~ \mu_2}^2}{\sigma_2}\bigg)\bigg ) \bigg ). \label{eq:9}
\end{equation}

The product of the copula density in Equation~\ref{eq:9}, the normal marginal of $logit(\pi_{i1}$) and $logit(\pi_{i2}$) in Equation~\ref{eq:8} form a bivariate normal distribution which characterize the model by \citet{Reitsma}, \citet{Arends}, \citet{Chu}, and \citet{Rileyb}, the so-called bivariate random-effects meta-analysis (BRMA) model, recommended as the appropriate method for meta-analysis of diagnostic accuracy studies. 
Study level covariate information explaining heterogeneity is introduced through the parameters of the marginal and the copula as follows 
\begin{equation}
\boldsymbol{\mu}_j = \textbf{X}_j\textbf{B}_j^{\top}. \label{eq:10}
\end{equation}								
\textbf{X}$_j$ is a \textit{n $\times$ p} matrix containing the covariates values for the mean sensitivity(\textit{j = 1}) and specificity(\textit{j = 2}). For simplicity, assume that \textbf{X}$_1$ = \textbf{X}$_2$ = \textbf{X}.   $\textbf{B}_j^\top$ is a \textit{p $\times$ 1} vector of regression parameters, and \textit{p} is the number of parameters.
By inverting the logit functions in Equation~\ref{eq:8}, we obtain
\begin{equation}\label{eq:11}
\pi_{ij} = logit^{-1} (\mu_j + \varepsilon_{ij}).
\end{equation} 									
Therefore, the meta-analytic sensitivity and specificity obtained by averaging over the random study effect, is given by, for \textit{j = 1, 2}
\begin{equation}\label{eq:12}
\E(\pi_{j} )~=~\E(logit^{-1} (\mu_j ~+~ \varepsilon_{ij}))
~=~\int_{-\infty}^{\infty}logit^{-1}(\mu_j ~+~ \varepsilon_{ij}) f(\varepsilon_{ij},~\sigma_j)~d\varepsilon_{ij},
\end{equation}								
assuming that $\sigma_1^2 > 0$ and $\sigma_2^2 > 0$. The integration in Equation~\ref{eq:12} has no analytical expression and therefore needs to be numerically approximated and the standard are not easily available. Using MCMC simulation in the Bayesian framework the meta-analytic estimates can be easily computed as well as a standard error estimate and a credible intervals $\E(\pi_{j})$ with minimum effort by generating predictions of the fitted bivariate normal distribution.

In the frequentist framework, it is more convenient however to use numerical averaging by sampling a large number \textit{M} of random-effects $\hat{\varepsilon}_{ij}$ from the fitted distribution and to estimate the meta-analytic sensitivity and specificity by \citep{Verbeke}, for \textit{j = 1, 2}
\begin{equation}\label{eq:13}
\hat{\E}(\pi_{j}) = \frac{1}{M} \sum_{i~ =~ 1}^{M}logit^{-1}(\hat{\mu}_j + \hat{\varepsilon}_{ij}).
\end{equation} 								
However, inference is not straightforward in the frequentist framework since the standard errors are not available. When $\varepsilon_{ij} = 0$, then
\begin{equation}\label{eq:14}
\E(\pi_{j} ~|~ \varepsilon_{ij} = 0) ~=~ logit^{-1}(\mu_j).
\end{equation}  						
Inference for $\E(\pi_{j} ~|~\varepsilon_{ij} ~=~ 0)$, as expressed in Equation~\ref{eq:14}, can be done in both Bayesian and frequentist framework.  The equation represents the mean sensitivity and specificity for a ``central" study with $\varepsilon_{ij} ~=~ 0$. Researchers often seem to confuse $\E(\pi_{j} ~|~\varepsilon_{ij} ~=~ 0)$ with $\E(\pi_{j})$ but due to the non-linear logit transformations, they are clearly not the same parameter. 

With the identity link function, no transformation on study-specific sensitivity and specificity is performed.  A natural choice for \textit{u} and \textit{v} would be beta distribution functions with parameters ($\alpha_1, ~\beta_1$) and ($\alpha_2, ~\beta_2$) respectively. Since $\pi_{ij} ~\sim ~ beta(\alpha_j, ~\beta_j)$, the meta-analytic sensitivity and specificity are analytically solved as follows
\begin{equation}\label{eq:15}
\E(\pi_{j} )
= \frac{\alpha_j}{\alpha_j + \beta_j},
\end{equation}

After reparameterising the beta distributions using the mean ($\mu_j ~=~ \frac{\alpha_j}{\alpha_j ~+~ \beta_j}$) and certainty ($\psi_j ~=~ \alpha_j ~+~ \beta_j$) or dispersion ($\varphi_j ~=~ \frac{1}{1~ +~ \alpha_j~+~\beta_j}$) parameters different link functions introduce covariate information to the mean, certainty/dispersion and association ($\rho$) parameters. A typical model parameterisation is \begin{align}
\boldsymbol{\mu}_j ~&=~ logit^{-1}(\textbf{XB}_j^{\top}), \nonumber \\
\boldsymbol{\psi}_j ~&=~ g(\textbf{WC}_j^{\top}), \nonumber \\
\boldsymbol{\alpha}_j ~&=~ \boldsymbol{\mu}_{j}  ~\circ~ \boldsymbol{\psi}_{j}, \nonumber \\
\boldsymbol{\beta}_j ~&=~ ( \textbf{1} ~-~ \boldsymbol{\mu}_{j})  ~\circ~ \boldsymbol{\psi}_{j}, \nonumber \\
\boldsymbol{\rho} ~&=~ tanh(\textbf{ZD}_j^{\top}) ~=~ \frac{exp(2 \times \textbf{ZD}_j^{\top}) ~-~ 1}{exp⁡(2\times \textbf{ZD}_j^{\top}) ~+~ 1}. \label{eq:16}
\end{align}								
\textbf{X, W} and \textbf{Z} are a \textit{n $\times$ p} matrices containing the covariates values for the mean, dispersion and correlation which we will assume has similar information and denoted by \textbf{X} for simplicity purpose, \textit{p} is the number of parameters,  $\textbf{B}_j^{\top}$ , $\textbf{V}_j^{\top}$ and $\textbf{D}_j^{\top}$ are a \textit{p $\times$ 1} vectors of regression parameters relating covariates to the mean, variance and correlation respectively. \textit{g}(.) is the log link to mapping $\textbf{XC}_j^{\top}$ to the positive real number line and $\circ$ is the Hadamard product. 

\subsubsection{Frank copula}

This flexible copula in the so-called family of Archimedean copulas was introduced by \citet{Frank}. The functional form of the copula and the density which is plugged in Equation~\ref{eq:3} is given by;
\begin{align}
C(F(\pi_{i1} ), ~F(\pi_{i2} ),\theta) ~&=~ - \frac{1}{\theta} ~log \bigg [ 1 + \frac{(e^{-\theta ~F(\pi_{i1})} - 1)
	(e^{-\theta~ F(\pi_{i2})} - 1)}{e^{-\theta} - 1} \bigg ],	\nonumber \\				
c(F(\pi_{i1} ), ~F(\pi_{i2} ),\theta) ~&=~  \frac {\theta~(1 - e^{-\theta})~e^{-\theta~(F(\pi_{i1}) ~ +~ F(\pi_{i2}))}}{[1 - e^{-\theta}-(1 - e^{-\theta ~F(\pi_{i1})})~(1 - e^{-\theta ~F(\pi_{i2})})]^2} . \label{eq:17}
\end{align}

Since $\theta \in \mathbb{R}$, both positive and negative correlation can be modelled, making this one of the more comprehensive copulas. When $\theta$ is 0, sensitivity and specificity are independent. For $\theta > 0$, sensitivity and specificity exhibit positive quadrant dependence and negative quadrant dependence when $\theta < 0$. The Spearman correlation $\rho_s$ and Kendall's tau $\tau_k$ can be expressed in terms of $\theta$ as
\begin{align}
\rho_s & = 1 - 12~\frac{D_2 (-\theta) ~-~ D_1(-\theta)}{\theta}, \nonumber \\
\tau_k & = 1 + 4~ \frac{D_1(\theta) ~-~ 1}{\theta}, \label{eq:18}
\end{align}										
where $D_j(\delta)$ is the Debye function defined as
\begin{equation}\label{eq:19}
D_j(\delta) ~=~ \frac{j}{\delta^j} \int_{\theta}^{\delta}\frac{t^j}{exp(t) ~-~ 1} ~dt, ~j~= 1, ~2.
\end{equation}

Covariate information is introduced in a similar manner as Equation~\ref{eq:16}. The identity link is used for the association parameter $\theta$. 
									
\subsubsection{Farlie-Gumbel-Morgenstern copula (FGM)}

This popular copula studied by \citet{Farlie}, \citet{Gumbel} and \citet{Morg} is defined as
\begin{align}
C(F(\pi_{i1} ), ~F(\pi_{i2}),~\theta) ~&= ~F(\pi_{i1})~F(\pi_{i2})[1 ~+~ \theta~(1 - F(\pi_{i1}))~(1 ~-~ F(\pi_{i2}))], \nonumber \\				
c(F(\pi_{i1} ), ~F(\pi_{i2}),~\theta) ~&=~ [1 ~+~ \theta~(2~F(\pi_{i1}) ~-~ 1)~(2~F(\pi_{i2}) ~-~ 1)].	\label{eq:21}
\end{align}				
Because $\theta \in (-1, 1)$, the Spearman correlation and Kendall's tau are expressed in terms of $\theta$ as $\theta/3$ and $2\theta/9$ respectively, making this copula only appropriate for data with weak dependence since $|\rho_s| ~\leq~ 1/3$. In a similar manner as in Equation~\ref{eq:16} the logit link, log/identity link and Fisher's z transformation can be used to introduce covariate information in modelling the mean, dispersion and association parameter.

\subsubsection{Clayton copula}

The Clayton copula function and density by \citet{Clayton} is defined as
\begin{align}
C(F(\pi_{i1}), ~F(\pi_{i2}),~\theta) &= [F(\pi_{i1})^{- \theta} ~+~ F(\pi_{i2})^{-\theta} ~-~ 1]^{\frac{- 1}{\theta}}, \nonumber \\
c(F(\pi_{i1}), ~F(\pi_{i2}),~\theta) &= (1 ~+~ \theta)~F(\pi_{i1})^{- (1 ~+~ \theta)} ~F(\pi_{i2})^{- (1 + \theta)}~[F(\pi_{i1} )^{- \theta}  + F(\pi_{i2})^{- \theta} ~-~ 1]^{\frac{- (2~\theta ~+~ 1)}{\theta}}. 	\label{eq:22}
\end{align}

Since $\theta \in (0, ~\infty)$, the Clayton copula typically models positive dependence; Kendall's tau equals $\theta/(\theta ~+~ 2)$. However, the copula function can be rotated by $90^\circ$ or $270^\circ$  to model negative dependence. The distribution and density functions following such rotations are given by
\begin{align}
C_{90} (F(\pi_{i1}), ~F(\pi_{i2}), ~\theta) ~&=~ F(\pi_{i2}) ~-~ C(1 ~-~ F(\pi_{i1}), ~F(\pi_{i2}), ~\theta), \nonumber \\
c_{90} (F(\pi_{i1}),~ F(\pi_{i2}),~\theta) ~&=~ (1 ~+~ \theta)(1 ~-~ F(\pi_{i1}))^{- (1 ~+~ \theta)}~F(\pi_{i2})^{- (1 ~+~ \theta)} ~[(1 - F(\pi_{i1}))^{-\theta} \nonumber \\ 
& +~ F(\pi_{i2})^{- \theta} - 1]^{\frac{- (2~\theta ~+~ 1)}{\theta}}, \label{eq:23}
\end{align}
and
\begin{align}
C_{270} (F(\pi_{i1}), ~F(\pi_{i2}), ~\theta) ~&= ~F(\pi_{i1})- C(F(\pi_{i1}), ~1 ~-~ F(\pi_{i2}),\theta),  \nonumber \\ 					
c_{270} (F(\pi_{i1}), ~F(\pi_{i2}), ~\theta) ~&=~(1 ~+~ \theta) ~F(\pi_{i1} )^{- (1 ~+~ \theta)}~ (1 ~-~ F(\pi_{i2} ))^{- (1 ~+~ \theta)}~[F(\pi_{i1})^{- \theta} \nonumber \\
&+ (1 ~-~ F(\pi_{i2}))^{- \theta} ~-~ 1]^{\frac{- (2~\theta ~+~ 1)}{\theta}}. \label{eq:24}
\end{align}											
The logit, log/identity and log/identity links can be used to introduce covariate information in modelling the mean ($\mu_j$), certainty ($\psi_j$)/dispersion ($\varphi_j$) and association ($\theta$) parameters respectively in the same way as in Equation~\ref{eq:16}.

Of course other copula functions that allow for negative association can be chosen. It is also an option to use known bivariate beta distributions. However, it is not always straightforward and analytically attractive to derive the corresponding copula function for all bivariate distributions. The use of existing bivariate beta distributions in meta-analysis of diagnostic accuracy studies has been limited because these densities model positive association ( e.g., \citet{Libby}, \citet{Olkina}), or both positive and negative association but over a restricted range( e.g., \citet{Sarmanov}). 

\section{Inference Framework and Software} \label{3}
Within the Bayesian framework, the analyst updates a prior opinion/information of a parameter based on the observed data whereas in the frequentist framework, the analyst investigates the behaviour of the parameter estimates in hypothetical repeated samples from a certain population. Due to its flexibility and use of MCMC simulations, complex modelling can often be implemented more easily within the Bayesian framework. By manipulating the prior distributions, Bayesian inference can circumvent identifiability problems whereas numerical approximation algorithms in frequentist inference without prior distributions can become stuck caused by identifiability problems. However, Bayesian methods typically require statistical expertise and patience because the MCMC simulations are computationally intensive. In contrast, most frequentist methods have been wrapped up in standard `procedures' that require less statistical knowledge and programming skills. Moreover frequentist methods are optimized with maximum likelihood estimation (MLE) that have much shorter run-times as opposed to MCMC simulations. \pkg{CopulaREMADA}~\citep{copularemada} is such a MLE based \proglang{R} package.

\subsection{The CopulaDTA package}
The \pkg{CopulaDTA} package is an extension of \pkg{rstan}~\citep{rstan}, the \proglang{R} interface to \proglang{Stan}~\citep{Stan} for diagnostic test accuracy data. \proglang{Stan} is a probabilistic programming language which has implemented Hamilton Monte Carlo(MHC) and uses No-U-Turn sampler (NUTS)~\citep{Gelman}. The package facilitates easy application of complex models and their visualization within the Bayesian framework with much shorter run-times.

\proglang{JAGS}~\citep{jags} is an alternative extensible general purpose sampling engine to \proglang{Stan}. Extending \proglang{JAGS} requires knowledge of \proglang{C++} to assemble a dynamic link library(DLL) module. From experience, configuring and building  the module is a daunting and tedious task especially in the Windows operation system. The above short-comings coupled with the fact that \proglang{Stan} tends to converge with fewer iterations even from bad initial values than \proglang{JAGS} made us prefer the \proglang{Stan} MCMC sampling engine.

The \pkg{CopulaDTA} package is available via the Comprehensive \proglang{R} Archive Network (CRAN) at \url{http://CRAN.R-project.org/package=CopulaDTA}. With a working internet connection, the \pkg{CopulaDTA} package is installed and loaded in \proglang{R} with the following commands
\begin{CodeInput}
R> install.packages("CopulaDTA", dependencies = TRUE)
	
R> library(CopulaDTA)	
\end{CodeInput}

The \pkg{CopulaDTA} package provide functions to fit bivariate beta-binomial distributions constructed as a product of two beta marginal distributions and copula densities discussed in Section~\ref{2}. The package also provides forest plots for a model with categorical covariates or with intercept only. Given the chosen copula function, a beta-binomial distribution is assembled up by the \code{cdtamodel} function which returns a \code{cdtamodel} object. The main function \code{fit} takes the \code{cdtamodel} object and fits the model to the given dataset and returns a \code{cdtafit} object for which \code{print}, \code{summary} and \code{plot} methods are provided for. 

\subsection{Model diagnostics}
To assess model convergence, mixing and stationarity of the chains, it is necessary to check the potential scale reduction factor $\hat{R}$, effective sample size (ESS), MCMC error and trace plots of the parameters. When all the chains reach the target posterior distribution, the estimated posterior variance is expected to be close to the within chain variance such that the ratio of the two, $\hat{R}$ is close to 1 indicating that the chains are stable, properly mixed and likely to have reached the target distribution. A large $\hat{R}$ indicates poor mixing and that more iterations are needed. Effective sample size indicates how much information one actually has about a certain parameter. When the samples are auto correlated, less information from the posterior distribution of our parameters is expected than would be if the samples were independent. ESS close to the total post-warm-up iterations is an indication of less autocorrelation and good mixing of the chains. Simulations with higher ESS have lower standard errors and more stable estimates. Since the posterior distribution is simulated there is a chance that the approximation is off by some amount; the Monte Carlo (MCMC) error. MCMC error close to 0 indicates that one is likely to have reached the target distribution.
 
\subsection[Model Comparison]{Model comparison and selection}
Watanabe-Alkaike Information Criterion (WAIC) \citep{Watanabe}, a recent model comparison tool to measure the predictive accuracy of the fitted models in the Bayesian framework, will be used to compare the models. WAIC can be viewed as an improvement of Deviance Information Criterion(DIC) which, though popular, is known to be have some problems \citep{Plummer}. WAIC is a fully Bayesian tool, closely approximates the Bayesian cross-validation, is invariant to reparameterisation and can be used for simple as well as hierarchical and mixture models.

\section[Data]{Datasets} \label{4}
\subsection{Telomerase data}
\citet{Glas} systematically reviewed the sensitivity and specificity of cytology and other markers including telomerase for primary diagnosis of bladder cancer. They fitted a bivariate normal distribution to the logit transformed sensitivity and specificity values across the studies allowing for heterogeneity between the studies. From the included 10 studies, they reported that telomerase had a sensitivity and specificity of 0.75 [0.66, 0.74] and 0.86 [0.71, 0.94] respectively. They concluded that telomerase was not sensitive enough to be recommended for daily use. This dataset is available within the package and the following commands
\begin{CodeInput}
R> data(telomerase)

R> telomerase
\end{CodeInput}	
loads the data into the R enviroment and generates the following output
\begin{Code}
   ID Dis TP NonDis  TN
1   1  33 25     26  25
2   2  21 17     14  11
3   3 104 88     47  31
4   4  26 16     83  80
5   5  57 40    138 137
6   6  47 38     30  24
7   7  42 23     12  12
8   8  33 27     20  18
9   9  17 14     32  29
10 10  44 37     29   7
\end{Code}
\code{ID} is the study identifier, \code{DIS} is the number of diseased, \code{TP} is the number of true positives, \code{NonDis} is the number of healthy and \code{TN} is the number of true negatives. 

\subsection{ASCUS triage data}
\citet{Arbyn} performed a Cochrane review on the accuracy of human papillomavirus testing and repeat cytology to triage of women with an equivocal Pap smear to diagnose cervical precancer. They fitted the BRMA model in \proglang{SAS} using \pkg{METADAS} on 10 studies where both tests were used. They reported absolute sensitivity of 0.909 [0.857, 0.944] and 0.715 [0.629, 0.788] for HC2 and repeat cytology respectively. The specificity was 0.607 [0.539, 0.68] and 0.684 [0.599, 0.758] for HC2 and repeat cytology respectively.  These data is used to demonstrate how the intercept only model is extended in a meta-regression setting. This dataset is also available within the package and the following commands
\begin{CodeInput}
R> data(ascus)

R> ascus
\end{CodeInput}	
loads the data into the R enviroment and generates the following output

\begin{Code}
   Test         StudyID  TP   FP  TN FN
1  RepC  Andersson 2005   6   14  28  4
2  RepC   Bergeron 2000   8   28  71  4
3  RepC Del Mistro 2010  20  191 483  7
4  RepC Kulasingam 2002  20   74 170  6
5  RepC     Lytwyn 2000   4   20  26  2
6  RepC      Manos 1999  48  324 570 15
7  RepC  Monsonego 2008  10   18 168 15
8  RepC      Morin 2001  14  126 214  5
9  RepC  Silverloo 2009  24   43 105 10
10 RepC    Solomon 2001 227 1132 914 40
11  HC2  Andersson 2005   6   17  25  4
12  HC2   Bergeron 2000  10   38  61  2
13  HC2 Del Mistro 2010  27  154 566  2
14  HC2 Kulasingam 2002  23  115 129  3
15  HC2     Lytwyn 2000   4   19  33  1
16  HC2      Manos 1999  58  326 582  7
17  HC2  Monsonego 2008  22  110  72  2
18  HC2      Morin 2001  17   88 253  2
19  HC2  Silverloo 2009  34   65  81  2
20  HC2    Solomon 2001 256 1050 984 11 
\end{Code}
\code{Test} is an explanatory variable showing the type of triage test, \code{StudyID} is the study identifier, \code{TP} is the number of true positives, \code{FP} is the number of false positives, \code{TN} is the number of true negatives, {FN} is the number of false negatives.

\section{The intercept only model} \label{5}
\begin{figure}[h]
	\fbox{\includegraphics[width=\textwidth, height=\textheight,keepaspectratio]{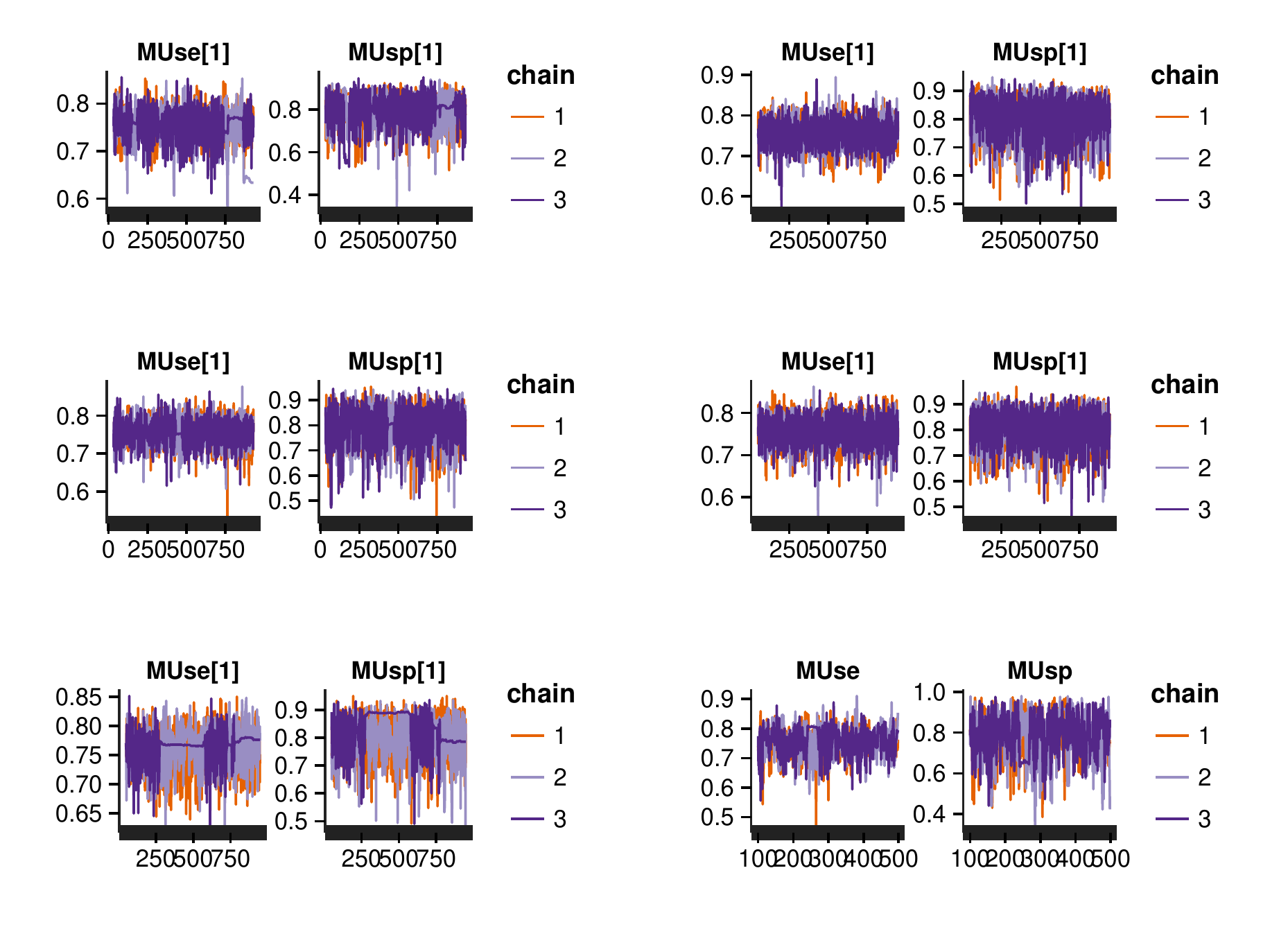}}
	\caption{Trace plots of the posterior mean sensitivity and specificity for the \code{telomerase} data as estimated by the Gaussian, Clayton 90$^\circ$ (C90) and 270$^\circ$ (C270), Farlie-Gumbel-Morgenstern (FGM) and Frank copula based bivariate beta and bivariate normal (BRMA) distributions.} 
	\label{Fig:1}
\end{figure}

The \pkg{CopulaDTA} package has five different correlation structures that result to five different bivariate beta-binomial distributions to fit to the data. The correlation structure is specified by indicating \code{copula ~=~"gauss"} or \code{"fgm"} or \code{"c90"} or \code{"270"} or \code{"frank"} in the \code{fitcopula} function. The Gaussian copula bivariate beta-binomial distribution is fitted to the \code{telomerase} data with the following code
\begin{CodeInput}
R> gauss.1 <- cdtamodel("gauss") 	

R> fitgauss.1 <- fit(
+		gauss.1,  
+		data = telomerase, 
+		SID = "ID", 
+		iter = 28000,
+		warmup = 1000,
+		thin = 30,
+		seed = 3)
\end{CodeInput}
By default, \code{chains = 3} and \code{cores = 3} and need not be specified unless otherwise. From the code above, \code{28000} samples are drawn from each of the \code{3} chains, the first \code{1000} samples are discarded and thereafter every \code{30}$^{th}$ draw kept such that each chain has 900 post-warm-up draws making a total of 2700 post-warm-up draws. The seed value, \code{seed = 3}, specifies a random number generator to allow reproducibility of the results and \code{cores = 3} allows for parallel-processing of the chains by using \code{3} cores, one core for each chain. They were no initial values specified and in that case, the program randomly generates random values satisfying the parameter constraints. 
The trace plots in the top-left panel of Figure~\ref{Fig:1} produced with the code below show satisfactory mixing of the chains and convergence.
\begin{CodeInput}
R> traceplot(fitgauss.1)
\end{CodeInput}

Next, obtain the model summary estimates as follows
\begin{CodeInput}
R> print(fitgauss.1, digits = 4)
\end{CodeInput}

\begin{Code}
Posterior marginal mean sensitivity and specificity
with 95

	Parameter    Mean   Lower   Upper n_eff  Rhat
MUse[1] Sensitivity  0.7540  0.6460  0.8119 122.4 1.010
MUsp[1] Specificity  0.8006  0.6235  0.9053 271.5 1.006
ktau[1] Correlation -0.8436 -0.9772 -0.3394 370.4 1.008

Model characteristics

Copula function: gauss, sampling algorithm: NUTS(diag_e)

Formula(1):  MUse ~ 1
Formula(2):  MUsp ~ 1
Formula(3):  Omega ~ 1
3 chain(s)each with iter=28000; warm-up=1000; thin=30.
post-warmup draws per chain=900;total post-warmup draws=2700.

Predictive accuracy of the model

Log point-wise predictive density (LPPD): -37.8529
Effective number of parameters: 7.0941
Watanabe-Akaike information Criterion (WAIC): 89.8940
\end{Code} 

From the output above, \code{n_eff} and \code{Rhat} both confirm proper mixing of the chains with little autocorrelation. The meta-analytic sensitivity \code{MUse[1]} and specificity \code{MUsp[1]} is 0.7540 [0.6460, 0.8119] and 0.8006 [0.6235, 0.9053] respectively. The Kendall's tau correlation between sensitivity and specificity is estimated to be -0.8436 [-0.9772, -0.3394]. 

\begin{figure}[h]
	\fbox{\includegraphics[width=\textwidth,height=\textheight,keepaspectratio]{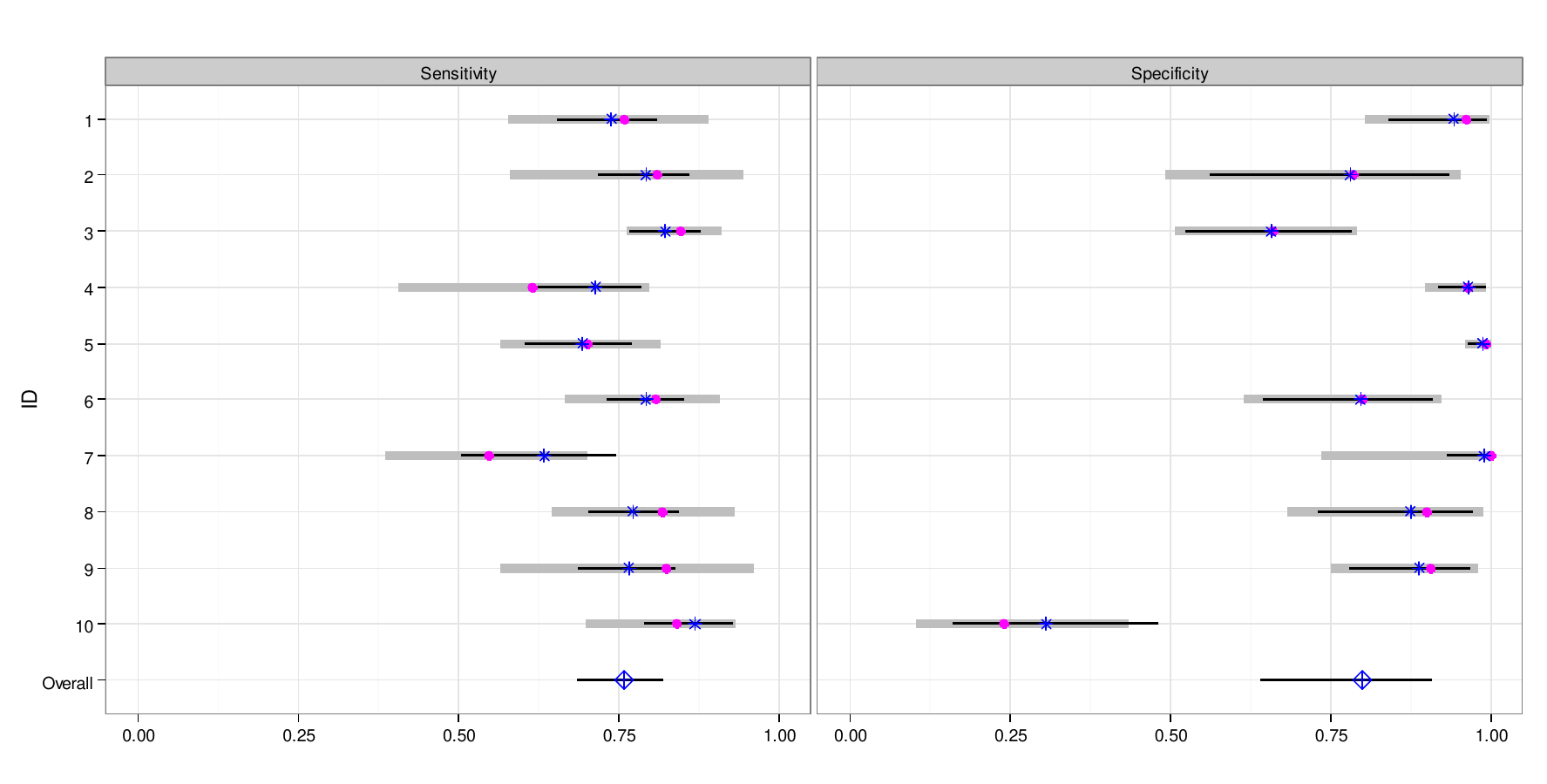}}
	\caption{Plot of the study-specific sensitivity and specificity (magenta points) and their corresponding 95 \% exact confidence intervals (thick grey lines), superimposed with the posterior estimates (blues stars) and their corresponding 95 \% credible intervals (think black lines). Posterior estimates from the Gaussian copula based bivariate beta distribution for the \code{telomerase} data.} 
	\label{Fig:2}
\end{figure}

The command below produces a forest plot in Figure~\ref{Fig:2}.
\begin{CodeInput}
R> plot(model1, graph = 3, title.3 = "" )
\end{CodeInput}

As observed in Figure~\ref{Fig:2}, the posterior study-specific sensitivity and specificity are less extreme and variable than the `observed' study-specific sensitivity and specificity. In other words, there is `shrinkage' towards the overall mean sensitivity and specificity as studies borrow strength from each other in the following manner: the posterior study-specific estimates depends on the global estimate and thus also on all other the studies.  

\begin{figure}[h]
	\fbox{\includegraphics[width=\textwidth,height=\textheight,keepaspectratio]{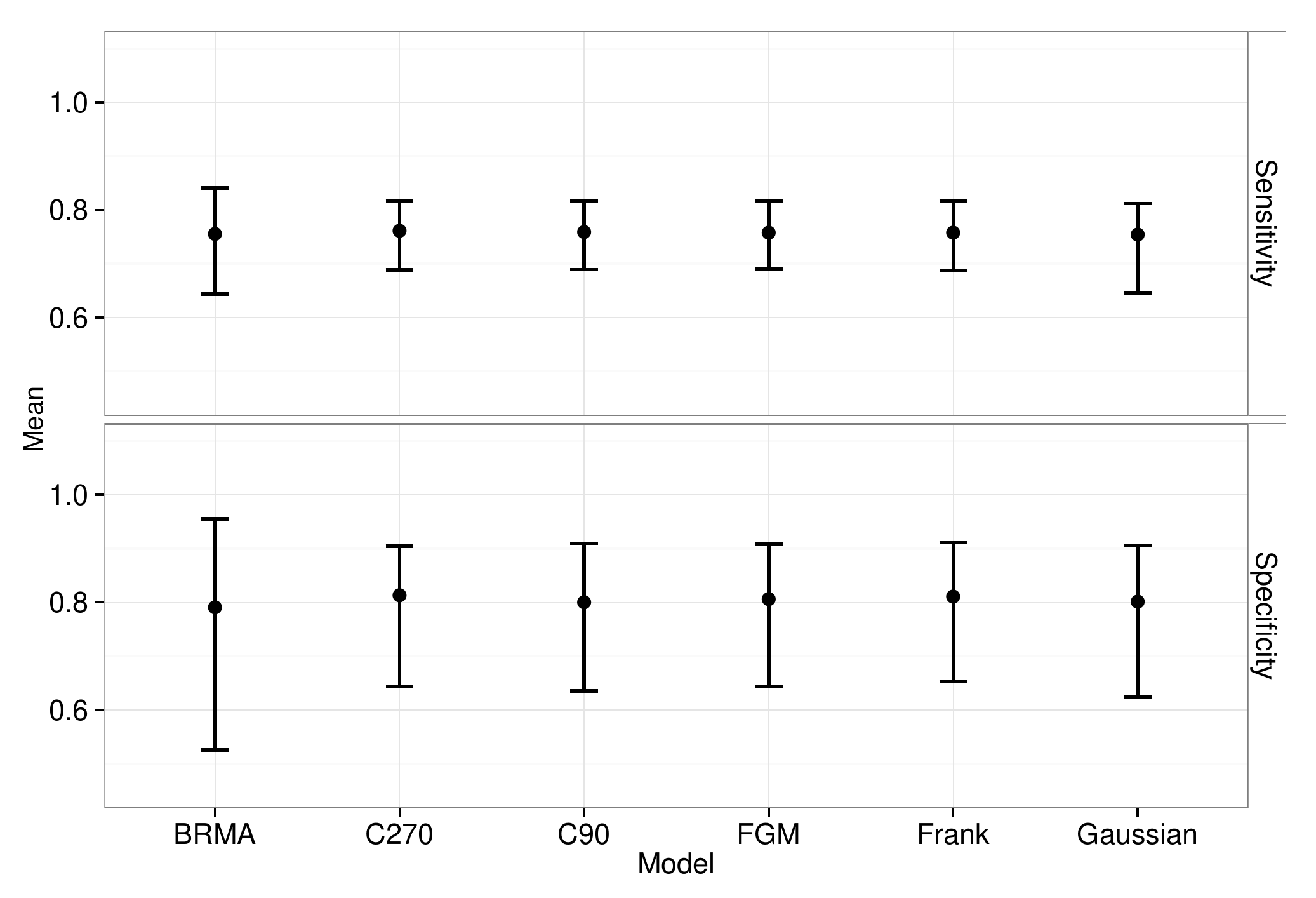}}
	\caption{Plot of the posterior meta-analytic sensitivity (upper) and specificity (lower) and the correspondinb 95\% credible intervals) as estimated by the Gaussian, Clayton 90$^\circ$ (C90) and 270$^\circ$ (C270), Farlie-Gumbel-Morgenstern (FGM) and Frank copula based bivariate beta and bivariate normal (BRMA) distributions for the \code{telomerase} data.} 
	\label{Fig:3}
\end{figure}

The mean sensitivity and specificity as estimated by the other four copula based bivariate beta distributions are in Table~\ref{Table:1} and graphically shown in Figure~\ref{Fig:3}. Though not presented here, the full code of the other four fitted copula based bivariate beta distributions is in the replication code. Figure~\ref{Fig:1} shows satisfactory chain mixing with little autocorrelation apart from the `Clayton270' model.  The Clayton copula is known to be unstable when the correlation parameter is close to the boundaries (-1 or 0) and this could be the reason why sampling from the posterior distribution was difficult.

For comparison purpose, the current recommended model; the BRMA, which uses normal marginals is also fitted to the data though it is not part of the \pkg{CopulaDTA} package. The model is first expressed in \proglang{Stan} modelling language in the code below and is stored within \proglang{R} environment as character string named \code{BRMA1}. 
\begin{CodeInput}
R> BRMA1 <- "
data{
	int<lower = 0> Ns;              
	int<lower = 0> tp[Ns];
	int<lower = 0>  dis[Ns];
	int<lower = 0>  tn[Ns];
	int<lower = 0>  nondis[Ns];
}
parameters{
	real etarho;                 
	vector[2] mul;               
	vector<lower = 0>[2] sigma; 
	vector[2] logitp[Ns];
	vector[2] logitphat[Ns]; 
}
transformed parameters{
	vector[Ns] p[2];
	vector[Ns] phat[2];            

	real MU[2];
	vector[2] mu;               
	real rho;					
	real ktau;                   
	matrix[2,2] Sigma; 
	
	rho <- tanh(etarho); 
	ktau <- (2/pi())*asin(rho);
	
	for (a in 1:2){
		for (b in 1:Ns){
			p[a][b] <- inv_logit(logitp[b][a]);
			phat[a][b] <- inv_logit(logitphat[b][a]);
			
		}
		mu[a] <- inv_logit(mul[a]);
	}
	
	MU[1] <- mean(phat[1]); 
	MU[2] <- mean(phat[2]); 
	
	Sigma[1, 1] <- sigma[1]^2;
	Sigma[1, 2] <- sigma[1]*sigma[2]*rho;
	Sigma[2, 1] <- sigma[1]*sigma[2]*rho;
	Sigma[2, 2] <- sigma[2]^2;
}
model{
	etarho ~ normal(0, 10);
	mul ~ normal(0, 10);
	sigma ~ cauchy(0, 2.5);
	
	for (i in 1:Ns){
		logitp[i] ~ multi_normal(mul, Sigma);
		logitphat[i] ~ multi_normal(mul, Sigma);
	}
	
	tp ~ binomial(dis,p[1]);
	tn ~ binomial(nondis, p[2]);
	
}
generated quantities{
	vector[Ns*2] loglik;
	
	for (i in 1:Ns){
		loglik[i] <- binomial_log(tp[i], dis[i], p[1][i]);
	}
	for (i in (Ns+1):(2*Ns)){
		loglik[i] <- binomial_log(tn[i-Ns], nondis[i-Ns], p[2][i-Ns]);
	}
}
"
\end{CodeInput}

Next, prepare the data by creating as list as follows
\begin{CodeInput}
R> datalist = list(
+	tp = telomerase$TP,
+	dis = telomerase$TP + telomerase$FN,
+	tn = telomerase$TN,
+	nondis = telomerase$TN + telomerase$FP,
+	Ns = 10)	
\end{CodeInput}

In the \code{data} block the dimensions and names of variables in the dataset are specified, here \code{Ns} indicate the number of studies in the dataset. The \code{parameters} block introduces the unknown parameters to be estimated. These are \code{etarho}; a scalar representing the Fisher's transformed form of the association parameter $\rho$, \code{mul};a \textit{2 $\times$ 1} vector representing the mean of sensitivity and specificity on the logit scale for a central study where the random-effect is zero, \code{sigma}; a \textit{2 $\times$ 1} vector representing the between study standard deviation of sensitivity and specificity on the logit scale, \code{logitp}; a \textit{Ns $\times$ 2} array of study-specific sensitivity in the first column and specificity in the second column on logit scale, and \code{logitphat}; a \textit{Ns $\times$ 2} array of predicted sensitivity in the first column and predicted specificity in the second column on logit scale.    

The parameters are further transformed in the \code{transformed parameters} block. Here, \code{p} is a \textit{2 $\times$ Ns} array of sensitivity in the first column and specificity in the second column after inverse logit transformation of \code{logitp}, and \code{phat} is a \textit{2 $\times$ Ns} array of predicted sensitivity in the first column and predicted specificity in the second column after inverse logit transformation of \code{logitphat} to be used in computing the meta-analytic sensitivity and specificity. \code{mu} is a \textit{2 $\times$ 1} vector representing the mean of sensitivity and specificity for a certain study with a random effect equal to 0, \code{MU} is a \textit{2 $\times$ 1} vector containing the meta-analytic sensitivity and specificity, \code{Sigma}; a \textit{2 $\times$ 2} matrix representing the variance-covarince matrix of sensitivity and specificity on the logit scale, \code{rho} and \code{ktau} are scalars representing the Pearson's and Kendall's tau correlation respectively. The prior distributions for the all parameters and data likelihood are defined in the \code{model} block.  Finally, in the \code{generated quantities} block, \code{loglik} is a \textit{(2Ns) $\times$ 1} vector of the log likelihood needed to compute the WAIC. 
 
Next, call the function \code{stan} from the \pkg{rstan} package to translate the code into \proglang{C++}, compile the code and draw samples from the posterior distribution as follows
\begin{CodeInput}
R> brma.1 <- stan(model_code = BRMA1,
+		data = datalist, 
+		chains = 3,
+		iter = 5000, 
+		warmup = 1000, 
+		thin = 10, 
+		seed = 3,
+		cores = 3)
\end{CodeInput}

The parameter estimates are extracted and the chain convergence and autocorrelation  examined further with the following code
\begin{CodeInput}
R> print(brma.1, pars = c('MU', 'mu', 'rho'), 
+	digits = 4, 
+	prob=c(0.025, 0.975))
\end{CodeInput}
The above lines of code print the following output
\begin{Code}
Inference for Stan model: d6a1713b5981968874c97152db2bb815.
3 chains, each with iter=5000; warmup=1000; thin=10; 
post-warmup draws per chain=400, total post-warmup draws=1200.

		mean se_mean     sd    2.5
MU[1]        0.7549  0.0020 0.0490  0.6438  0.8408   594 0.9995
MU[2]        0.7901  0.0056 0.1121  0.5252  0.9554   397 1.0036
mu[1]        0.7681  0.0014 0.0367  0.6850  0.8425   714 0.9999
mu[2]        0.8971  0.0026 0.0715  0.7189  0.9822   763 1.0058
rho         -0.9338  0.0092 0.1224 -0.9993 -0.5711   175 1.0208

Samples were drawn using NUTS(diag_e) at Mon Dec 07 14:59:26 2015.
For each parameter, n_eff is a crude measure of effective sample size,
and Rhat is the potential scale reduction factor on split chains (at 
convergence, Rhat=1).
\end{Code}

The meta-analytic sensitivity (\code{MU[1]}) and specificity (\code{MU[2]}) and 95\% credible intervals are 0.7549[0.6438, 0.8408] and 0.7901[0.5252, 0.9554] respectively. This differs from what the authors published (0.75[0.66, 0.74] and 0.86[0.71, 0.94]) in two ways. The authors fitted the standard bivariate normal distribution to the logit transformed sensitivity and specificity values across the studies allowing for heterogeneity between the studies as expressed in Equation~\ref{eq:6} and disregarded the higher level of the hierarchical model expressed in Equation~\ref{eq:5}. Because of this the authors had to use a continuity correction of 0.5 since the seventh study had `observed' specificity equal to 1, a problem not encountered in the hierarchical model. Secondly the authors do not report the meta-analytic values but rather report the mean sensitivity(\code{mu[1]}) and specificity (\code{mu[2]}) for a particular, hypothetical study with random-effect equal to zero, which in our case is  0.7681[0.6850, 0.8425] and 0.8971[0.7189, 0.98227] respectively and is comparable to what the authors reported. This discrepancy between \code{MU} and \code{mu} will indeed increase with increase in the between study variability.

\begin{table}[h]
	\centering
	\begin{tabular}{|c|l|c|c|c|c|c|c|}
		\hline
		\multicolumn{1}{|l|}{Model} & Parameter   & \multicolumn{1}{l|}{Mean} & \multicolumn{1}{l|}{Lower} & \multicolumn{1}{l|}{Upper} & \multicolumn{1}{l|}{n\_eff} & \multicolumn{1}{l|}{Rhat} & \multicolumn{1}{l|}{WAIC} \\ \hline
		\multirow{3}{*}{Gaussian}   & Sensitivity & 0.7540                    & 0.6460                     & 0.8119                     & 122                         & 1.0076                    & \multirow{3}{*}{89.8940}  \\ \cline{2-7}
		& Specificity & 0.8006                    & 0.6235                     & 0.9053                     & 271                         & 1.0061                    &                           \\ \cline{2-7}
		& Correlation & -0.8436                   & -0.9772                    & -0.3394                    & 370                         & 1.0083                    &                           \\ \hline
		\multirow{3}{*}{C90}        & Sensitivity & 0.7579                    & 0.6894                     & 0.8165                     & 2046                        & 1.0008                    & \multirow{3}{*}{92.4859}  \\ \cline{2-7}
		& Specificity & 0.7996                    & 0.6352                     & 0.9100                     & 1473                        & 1.0053                    &                           \\ \cline{2-7}
		& Correlation & -0.7338                   & -0.9830                    & 0.0000                     & 1347                        & 1.0000                    &                           \\ \hline
		\multirow{3}{*}{C270}       & Sensitivity & 0.7606                    & 0.6888                     & 0.8165                     & 807                         & 1.0097                    & \multirow{3}{*}{90.5935}  \\ \cline{2-7}
		& Specificity & 0.8125                    & 0.6439                     & 0.9048                     & 61                          & 1.0311                    &                           \\ \cline{2-7}
		& Correlation & -0.7526                   & -0.9800                    & 0.0000                     & 17                          & 1.0601                    &                           \\ \hline
		\multirow{3}{*}{FGM}        & Sensitivity & 0.7576                    & 0.6901                     & 0.8169                     & 2700                        & 0.9994                    & \multirow{3}{*}{95.4854}  \\ \cline{2-7}
		& Specificity & 0.8053                    & 0.6427                     & 0.9085                     & 2700                        & 1.0001                    &                           \\ \cline{2-7}
		& Correlation & -0.1871                   & -0.2222                    & 0.2222                     & 2377                        & 1.0012                    &                           \\ \hline
		\multirow{3}{*}{Frank}      & Sensitivity & 0.7576                    & 0.6880                     & 0.8165                     & 2693                        & 1.0008                    & \multirow{3}{*}{89.9743}  \\ \cline{2-7}
		& Specificity & 0.8097                    & 0.6526                     & 0.9112                     & 2582                        & 1.0007                    &                           \\ \cline{2-7}
		& Correlation & -0.7073                   & -0.8504                    & -0.1852                    & 2700                        & NA                        &                           \\ \hline
		\multirow{3}{*}{BRMA}       & Sensitivity & 0.7549                    & 0.6438                     & 0.8408                     & 594                         & 0.9995                    & \multirow{3}{*}{86.6359}  \\ \cline{2-7}
		& Specificity & 0.7901                    & 0.5252                     & 0.9554                     & 397                         & 1.0036                    &                           \\ \cline{2-7}
		& Correlation & -0.8204                   & -0.9755                    & -0.3870                    & 39                          & 1.0556                    &                           \\ \hline
	\end{tabular}
	\caption{The posterior mean, 95\% credible interval, effective sample size and potential scale reduction factor $\hat{R}$ factor for the marginal means and correlation parameters as estimated by the Gaussian, Clayton 90$^\circ$ (C90) and 270$^\circ$ (C270), Farlie-Gumbel-Morgenstern (FGM) and Frank copula based bivariate beta and bivariate normal (BRMA) distributions for the \code{telomerase} data. }
	\label{Table:1}
\end{table}

\subsection{Model comparison}
Table~\ref{Table:1} shows that the correlation as estimated by the BRMA model and the Gaussian copula bivariate beta  are more extreme but comparable to the estimates from the Frank,  $90^{\circ}$- and $270^{\circ}$- Clayton copula. On the other extreme is the estimate from the model  FGM copula bivariate beta and this is due to the constraints on the association parameter in the FGM copula where values lie within |2/9|.

In Figure~\ref{Fig:3}, the marginal mean sensitivity and specificity from the five bivariate beta distributions are comparable with subtle  differences in the 95 percent credible intervals despite differences in the correlation structure. 

\citet{Glas} and \citet{Rileyb} estimated the Pearson's correlation parameter in the BRMA model $\rho$ as -1 within the frequentist framework. Using maximum likelihood estimation, \citet{Rileya} showed that the between-study correlation from the BRMA is often estimated as +/-1. Without estimation difficulties, Table \ref{Table:1} shows an estimated Pearson's correlation of -0.9338[-0.9993, -0.57118]. This is because Bayesian methods are not influenced by sample size and therefore able to handle cases of small sample sizes with less issues. 

Essentially, all the six models are equivalent in the first level of hierarchy and differ in specifying the prior distributions for the `study-specific' sensitivity and specificity. As thus, the models should have the same number of parameters in which case it makes sense then to compare the log predictive densities. Upon inspection, the log predictive densities from the six models are practically equivalent (min=37.40, max=38.62) but the effective number of parameters differed a bit (max=5.9, max=9.1). Apparently, the last column of Table \ref{Table:1} indicates that the BRMA fits the data best based on the WAIC. 

\section{Meta-regression} \label{6}
The \code{ascus} dataset has \code{Test} as a covariate. The covariate is used as it is of interest to study its effect on the joint distribution of sensitivity and specificity (including the correlation). The following code fits the FGM copula based bivariate beta-binomial distribution to the data
\begin{CodeInput}
R> fgm.2 <- cdtamodel(copula = "fgm",
+		 modelargs = list(formula.se = StudyID ~ Test + 0))
	
R> fitfgm.2 <- fit(fgm.2,
+		data = ascus, 
+		SID = "StudyID",  
+		iter = 19000,
+		warmup = 1000,
+		thin = 20,
+		seed = 3)
\end{CodeInput}

\begin{figure}[h]
	\fbox{\includegraphics[width=\textwidth,height=\textheight,keepaspectratio]{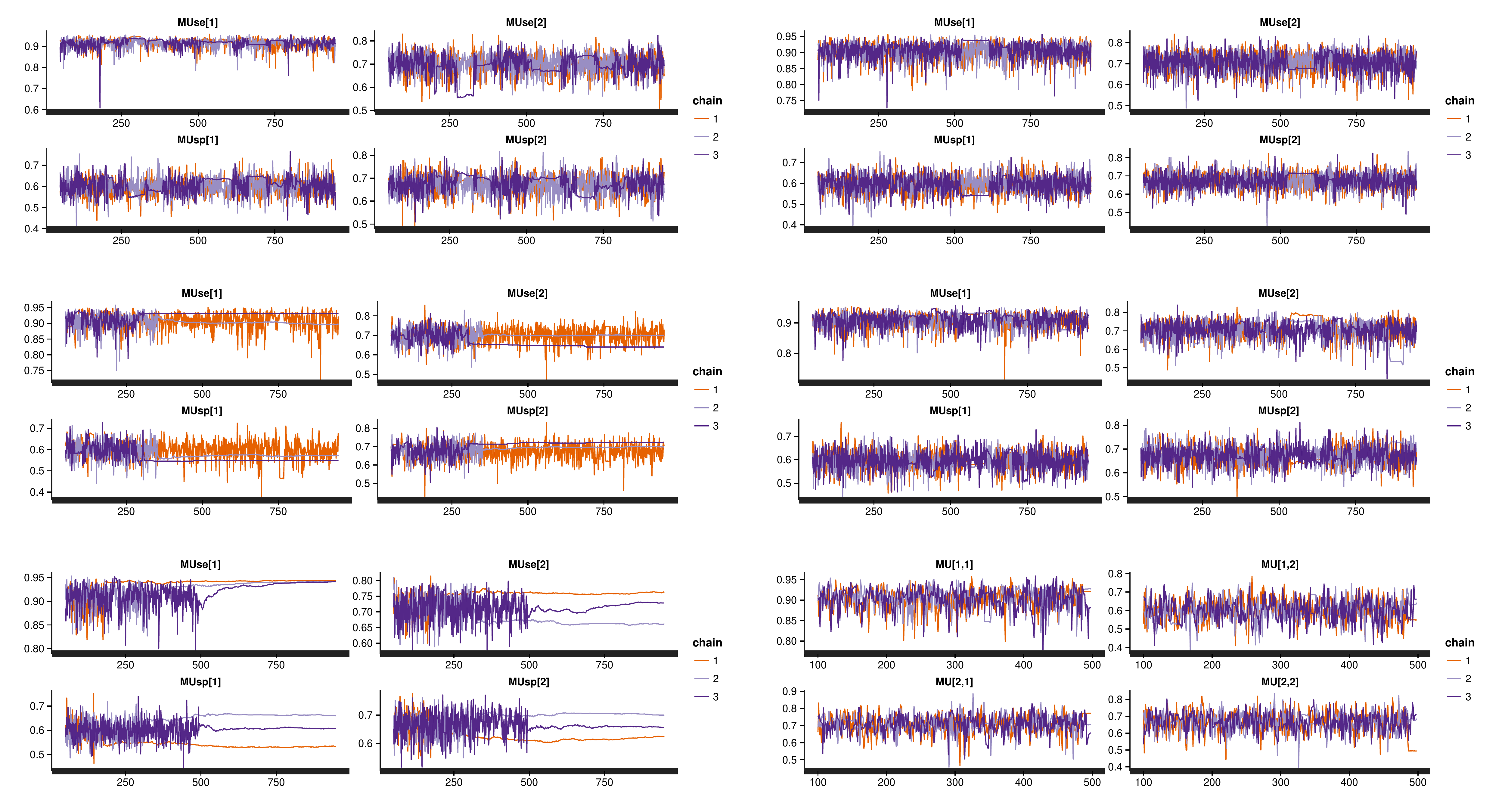}}
		\caption{Trace plots of the posterior mean sensitivities and specificities for the \code{ascus} data as estimated by the Gaussian, Clayton 90$^\circ$ (C90) and 270$^\circ$ (C270), Farlie-Gumbel-Morgenstern (FGM) and Frank copula based bivariate beta and bivariate normal (BRMA) distributions.} 
	\label{Fig:4}
\end{figure}

\begin{figure}[h]
	\fbox{\includegraphics[width=\textwidth,height=\textheight,keepaspectratio]{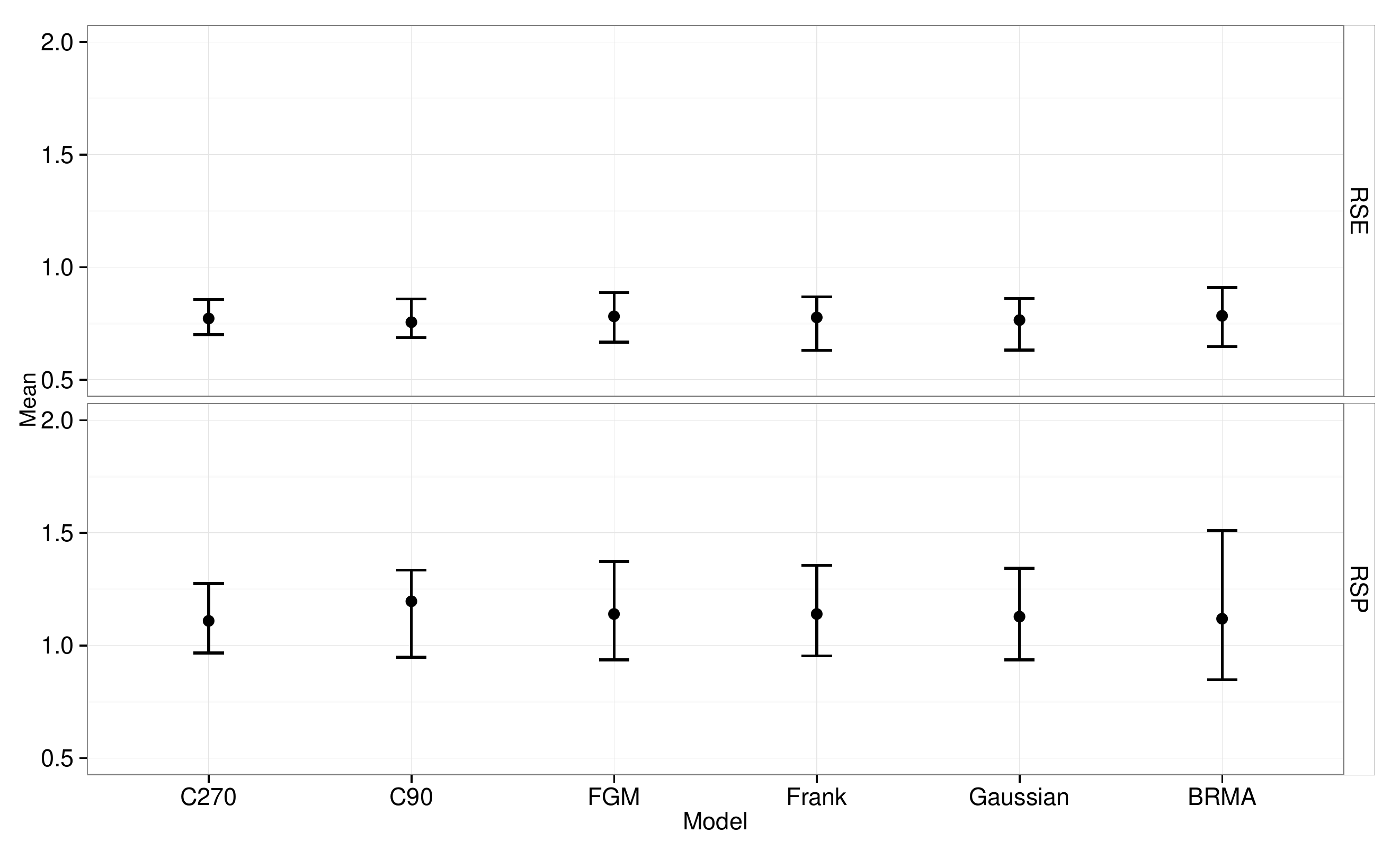}}
	\caption{Pooled relative sensitivity (on top) and relative specificity (bottom) of repeat cytology (posterior mean and 95\% credible intervals) compared to HPV testing with HC2 to detect cervical precancer in women with an atypical Pap smear estimated.}
	\label{Fig:5}
\end{figure}

Figure~\ref{Fig:4} shows the trace plots for all the six models fitted to the \code{ascus} data where all parameters, including the correlation parameter(except the BRMA) are modeled as a function of the covariate. There is proper chains mixing and convergence except for the case of the Clayton copula based bivariate beta. From the posterior relative sensitivity and specificity plotted in Figure~\ref{Fig:5}, all the models that converged generally agree that repeat cytology was less sensitive than HC2 without significant loss in specificity.

\begin{table}[h]
	\centering
	\begin{tabular}{|l|l|c|c|c|c|c|c|}
		\hline
		Model                     & Test & \multicolumn{1}{l|}{Mean} & \multicolumn{1}{l|}{Lower} & \multicolumn{1}{l|}{Upper} & \multicolumn{1}{l|}{n\_eff} & \multicolumn{1}{l|}{Rhat} & \multicolumn{1}{l|}{WAIC}  \\ \hline
		\multirow{2}{*}{Gaussian} & HC2  & -0.4799                   & -0.9902                    & 0.8972                     & 403                         & 1.0027                    & \multirow{2}{*}{5350.6348} \\ \cline{2-7}
		& Repc & -0.9164                   & -0.9972                    & -0.6122                    & 390                         & 1.0004                    &                            \\ \hline
		\multirow{2}{*}{C90}      & HC2  & -0.1122                   & -0.9238                    & 0.0000                     & 14                          & 1.0910                    & \multirow{2}{*}{5347.9687} \\ \cline{2-7}
		& Repc & -0.8554                   & -0.9825                    & -0.3763                    & 38                          & 1.0553                    &                            \\ \hline
		\multirow{2}{*}{C270}     & HC2  & -0.0524                   & -0.8139                    & 0.0000                     & 43                          & 1.0926                    & \multirow{2}{*}{5339.1644} \\ \cline{2-7}
		& Repc & -0.7898                   & -0.9783                    & -0.3691                    & 5                           & 1.6967                    &                            \\ \hline
		\multirow{2}{*}{FGM}      & HC2  & -0.0836                   & -0.2222                    & 0.2222                     & 451                         & 1.0072                    & \multirow{2}{*}{5355.6485} \\ \cline{2-7}
		& Repc & -0.1999                   & -0.2222                    & 0.1704                     & 2584                        & 0.9994                    &                            \\ \hline
		\multirow{2}{*}{Frank}    & HC2  & -0.5238                   & -0.8201                    & 0.6067                     & 2700                        & NA                        & \multirow{2}{*}{5353.1396} \\ \cline{2-7}
		& Repc & -0.7410                   & -0.8627                    & -0.3253                    & 2700                        & NA                        &                            \\ \hline
		BRMA                      & Both & -0.8483                   & -0.9954                    & -0.4391                    & 5                           & 1.1967                    & 5348.3116                  \\ \hline
	\end{tabular}
	\caption{The posterior mean, 95\% credible intervals, effective sample size, potential scale reduction factor of the correlation parameter(s)  as estimated by the Gaussian, Clayton 90$^\circ$ (C90) and 270$^\circ$ (C270), Farlie-Gumbel-Morgenstern (FGM) and Frank copula based bivariate beta and bivariate normal (BRMA) distributions for the \code{ascus} dataset.}
	\label{Table:2}
\end{table}

The \code{n\_eff} in Table~\ref{Table:2} indicate substantial autocorrelation in sampling the correlation parameters except in the `Gaussian', `FGM' and `Frank' models. From the copula based bivariate beta distributions, it is apparent that the correlation between sensitivity and specificity in HC2 and repeat cytology is different.

The `Clayton' models have the lowest WAIC even though sampling from the posterior distribution was difficult as seen in their trace plots in Figure~\ref{Fig:4} and the \code{n_eff} and \code{Rhat} in Table~\ref{Table:2}. The difficulty in sampling from the posterior could be signalling over-parameterisation of the correlation structure. It would thus be interesting to re-fit the models using only one correlation parameter and compare the models.
WAIC is known to fail in certain settings and this examples shows that it is crucial to check the adequacy of the fit and plausibility of the model and not blindly rely on an information criterion to select the best fit to the data.

\section[Discussion]{Discussion} \label{7}
 Copula-based models offer great flexibility and ease but their use is not without caution. While the copulas used in this paper are attractive as they are mathematically tractable, \citet{Mikosch} and \citet{Genest} noted that it might be difficult to estimate copulas from data. Furthermore, the concepts behind copula models is slightly more complex and therefore require statistical expertise to understand and program them as they are not yet available as standard procedure/programs in statistical software. 
 
 In this paper, several advanced statistical models for meta-analysis of diagnostic accuracy studies were briefly discussed.  The use of the \proglang{R}  package \pkg{CopulaDTA} within the flexible \pkg{Stan} interface was demonstrated and shows how complex models can be implemented in a convenient way.

In most practical situations, the marginal mean structure is of primary interest and the correlation structure is treated a nuisance making the choice of copula less critical. Nonetheless, an appropriate correlation structure is critical in the interpretation of the random variation in the data as well as obtaining valid model-based inference for the mean structure.

When the model for the mean is correct but the true distribution is misspecified, the estimates of the model parameters will be consistent but the standard errors will be incorrect \cite{Agresti}. Nonetheless, the bivariate beta distribution has the advantage to allow direct joint modelling of sensitivity and specificity, without the need of any transformation, and consequently providing estimates with the appropriate meta-analytic interpretation but with the disadvantage of being more computationally intensive for some of the copula functions.

\citet{Leeflang} showed that the sensitivity and specificity often vary with disease prevalence. The models presented above can easily be extended and implemented to jointly model prevalence, sensitivity and specificity using tri-variate copulas. 

There were some differences between the models in estimating the meta-analytic sensitivity and specificity and the correlation. Therefore, further research is necessary to investigate the effect of certain parameters, such as the number of studies, sample sizes and misspecification of the joint distribution on the meta-analytic estimates.

\section[Conclusion]{Conclusion} \label{8}
The proposed Bayesian joint model using copulas to construct bivariate beta distributions, provides estimates with both the appropriate marginal as well as conditional interpretation, as  opposed to the typical BRMA model which estimates sensitivity and specificity for specific studies with a particular value for the random-effects.  Furthermore, the models do not have estimation difficulties with small sample sizes or large between-study variance because: \textbf{i}) the between-study variances are not constant but depends on the underlying means and \textbf{ii} Bayesian methods are less influenced by small samples sizes.

The fitted models generally agree that the mean specificity was slightly lower than what \citet{Glas} reported and based on this we conclude that telomerase was not sensitive and specific enough to be recommended for daily use.

In the ASCUS triage data, conclusion based on the fitted models is in line with what the authors conclude: that HC2 was considerably more sensitive but sligthly and non-significantly less specific than repeat cytology to triage women with an equivocal Pap smear to diagnose cervical precancer.

While the BRMA had the lowest WAIC for both datasets,  we still recommend modelling of sensitivity and specificity using bivariate beta distributions as they easily and directly provide meta-analytic estimates. 

\subsection*{Competing interests}
The authors declare that they have no competing interests.
\subsection*{Author's contributions}
M. Arbyn designed the OPSADAC project (Optimisation of statistical procedures to assess the diagnostic accuracy of cervical cancer screening tests) of which this study is a part of. Victoria and M. Aerts conceptualized and initiated the study. Victoria wrote the code, analysed the data and drafted manuscript. M. Arbyn and M. Aerts edited the manuscript. All authors reviewed and approved the final manuscript.

\subsection*{Acknowledgements}
V. Nyaga recieved financial support from the Scientific Institute of Public Health (Brussels) through the OPSADAC project. M. Arbyn was supported by the COHEAHR project funded by the 7th Framework Programme of the European Commission (grant No 603019). M. Aerts was supported by the IAP research network nr P7/06 of the Belgian Government (Belgian Science Policy).

\end{document}